\documentclass[aps,prl,twocolumn,english,superscriptaddress,10pt]{revtex4-1}
\usepackage{amstext,amssymb,graphicx,babel,soul,upgreek,bm}

\usepackage[usenames, dvipsnames]{color}

\usepackage{comment}

\def\RM#1 {\textbf{\color{red} [RM: {#1}] }}
\def\JE#1 {\textbf{\color{red} [JE: {#1}] }}
\def\RT#1 {\textbf{\color{red} [RT: {#1}] }}
\def\MG#1 {\textbf{\color{red} [MG: {#1}] }}
\def\IM#1 {\textbf{\color{red} [IM: {#1}] }}

\def\TW#1 {\textbf{\color{red} [TW: {#1}] }}

\def\eqref#1{(\ref{#1})}

\def\be{\begin{equation}}
\def\ee{\end{equation}}
\def\bea{\begin{eqnarray}}
\def\eea{\end{eqnarray}}

\begin{document}

\title{Turbulent hydrodynamics in strongly correlated Kagome metals}

\author{Domenico Di Sante}
\affiliation{Institut f\"ur Theoretische Physik und Astrophysik and W\"urzburg-Dresden Cluster of Excellence ct.qmat, Julius-Maximilians-Universit\"at W\"urzburg, Am Hubland 97074 W\"{u}rzburg, Germany}

\author{Johanna Erdmenger}
\affiliation{Institut f\"ur Theoretische Physik und Astrophysik and W\"urzburg-Dresden Cluster of Excellence ct.qmat, Julius-Maximilians-Universit\"at W\"urzburg, Am Hubland 97074 W\"{u}rzburg, Germany}

\author{Martin Greiter}
\affiliation{Institut f\"ur Theoretische Physik und Astrophysik and W\"urzburg-Dresden Cluster of Excellence ct.qmat, Julius-Maximilians-Universit\"at W\"urzburg, Am Hubland 97074 W\"{u}rzburg, Germany}

\author{Ioannis Matthaiakakis}
\affiliation{Institut f\"ur Theoretische Physik und Astrophysik and W\"urzburg-Dresden Cluster of Excellence ct.qmat, Julius-Maximilians-Universit\"at W\"urzburg, Am Hubland 97074 W\"{u}rzburg, Germany}

\author{Ren\'e Meyer}
\affiliation{Institut f\"ur Theoretische Physik und Astrophysik and W\"urzburg-Dresden Cluster of Excellence ct.qmat, Julius-Maximilians-Universit\"at W\"urzburg, Am Hubland 97074 W\"{u}rzburg, Germany}

\author{David Rodr\'iguez Fern\'andez}
\affiliation{Institut f\"ur Theoretische Physik und Astrophysik and W\"urzburg-Dresden Cluster of Excellence ct.qmat, Julius-Maximilians-Universit\"at W\"urzburg, Am Hubland 97074 W\"{u}rzburg, Germany}

\author{Ronny Thomale}
\affiliation{Institut f\"ur Theoretische Physik und Astrophysik and W\"urzburg-Dresden Cluster of Excellence ct.qmat, Julius-Maximilians-Universit\"at W\"urzburg, Am Hubland 97074 W\"{u}rzburg, Germany}

\author{Erik van Loon}
\affiliation{Institut f\"ur Theoretische Physik, Universit\"at Bremen, Otto-Hahn-Allee 1, 28359 Bremen, Germany}

\author{Tim Wehling}
\affiliation{Institut f\"ur Theoretische Physik, Universit\"at Bremen, Otto-Hahn-Allee 1, 28359 Bremen, Germany}

\maketitle

\textbf{
A current challenge in condensed matter physics is the realization of strongly correlated, viscous electron fluids~\cite{polini2019viscous}.  These fluids are not amenable to the perturbative methods of Fermi liquid theory, but can be described by holography, that is, by mapping them onto a weakly curved gravitational theory via gauge/gravity duality~\cite{Policastro:2001yc,Buchel2004,Kovtun2005}.  The canonical system considered for realizations has been graphene, which possesses Dirac dispersions at low energies as well as significant Coulomb interactions between the electrons~\cite{novoselov2005two}.  In this work, we show that Kagome systems with electron fillings adjusted to the Dirac nodes of their band structure~\cite{sc-herbert} provide a much more compelling platform for realizations of viscous electron fluids, including non-linear effects such as turbulence~\cite{Landau:1987}.  In particular, we find that in stoichiometric Scandium (Sc) Herbertsmithite, the fine-structure constant, which measures the effective Coulomb interaction and hence reflects the strength of the correlations, is enhanced by a factor of about 3.2 as compared to graphene, due to orbital hybridization.  
We employ ho\-lo\-graphy to estimate the ratio of the shear viscosity over the entropy density in Sc-Herbertsmithite, and find it about three times smaller than in graphene.  These findings put, for the first time, the turbulent flow regime described by holography \cite{adams2014holographic} within the reach of experiments. 
}

Electrons in solids typically interact not only with impurities and phonons, but also with each other via the Coulomb interaction. If the momentum relaxing effects of impurities and phonons are weak, the Coulomb interaction can become dominant and lead to local thermalization and the formation of an electronic fluid. Thus, the length and time scales over which thermalization occurs are controlled by the strength of the Coulomb interaction. The regime of electron hydrodynamics has been realized in several systems~\cite{Molenkamp:1994ii,Molenkamp:1994kb,Moll:2016ju,gooth2017electrical,Bandurin2018}, giving rise to new transport properties~\cite{Torre:2015eja,levitov2016electron,bandurin2016negative,Pellegrino:2016kp,kumar2017superballistic} distinctly different from the ballistic regime. Therefore, characterizing the transport properties of viscous electronic fluids is tantamount to determining the strength of the Coulomb coupling  $\alpha$. 

\begin{figure}[!t]
\centering
\includegraphics[width=\columnwidth]{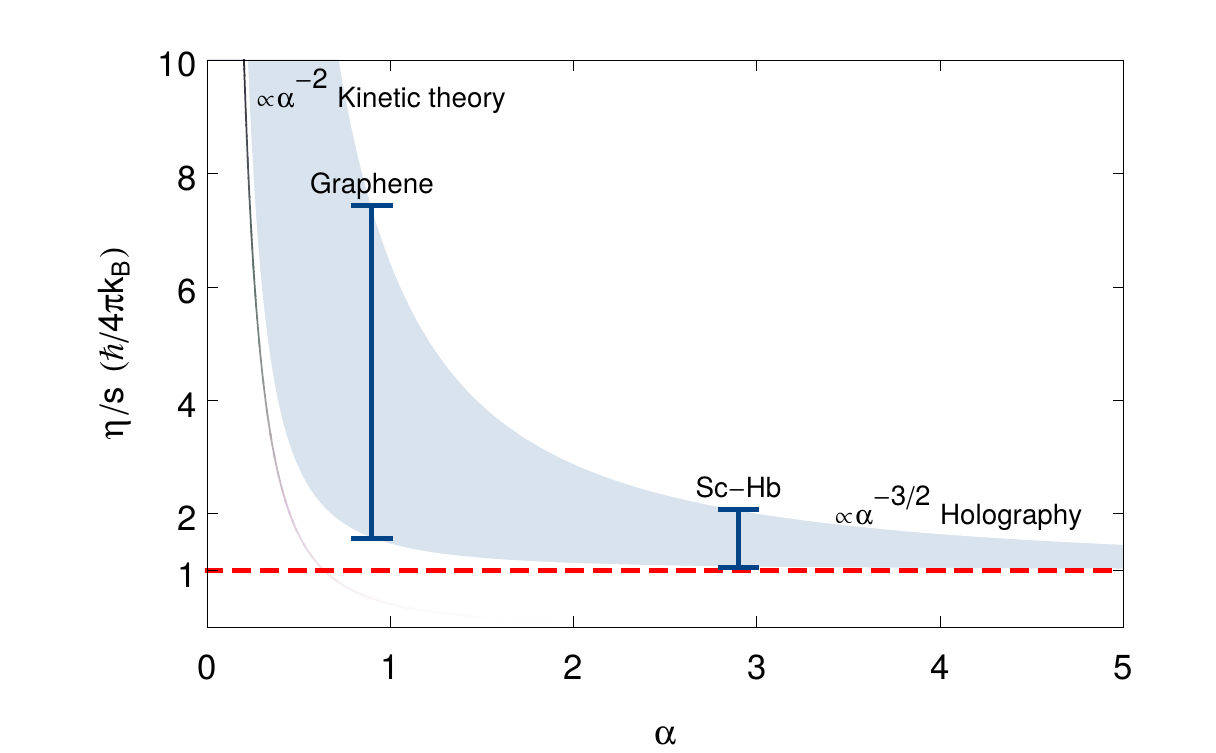}
\caption{
{\bf $\boldsymbol{\eta}/\boldsymbol{s}$ as a function of the coupling
strength.} Black line: Prediction in the weak coupling regime,
$\left(\eta/s\right)\propto \alpha^{-2}$, reliable for small values of
$\alpha$. The red dashed line corresponds to the universal holographic
value, $\eta/s = \hbar/4\pi k_B$. The blue shaded region, for which
$\eta/s$ is given by equation~\eqref{eq:FiniteCouplingCorrectionsMain},
represents the holographic prediction. This shaded region parametrizes
the class of extrapolating models beyond the $\alpha \rightarrow \infty$
limit (see Methods). Notice that Sc-Herbertsmithite (Sc-Hb) shows a much smaller variance
than graphene (given by the vertical blue bars at $\alpha = 2.9$ for Sc-Hb and
$\alpha = 0.9$ as a representative value for graphene, respectively), providing further support for the
applicability of holographic methods to correlated Kagome metals.
}
\label{fig:eta/s}
\end{figure}

The Coulomb interaction also mediates energy and momentum transfer between the thermalized regions of the fluid, and thereby controls transport coefficients such as the shear viscosity $\eta$. Since direct access to the Coulomb coupling $\alpha$ proves difficult, we focus in this work on the easily accessible shear viscosity, or more precisely on the ratio between $\eta$ and the entropy density of the fluid, $s$. As explained in Methods, the shear viscosity $\eta$ is straightforwardly obtained from a Kubo formula (see equation~\eqref{eq:eta over s} in Methods), and the entropy density $s$ from thermodynamics (see  equation~\eqref{eq:GibbsDuhem} in Methods). For a Dirac fluid, the ratio $\eta/s$ is equal to the temperature multiplied by the kinematic viscosity $\nu$, which is experimentally accessible~\cite{Pellegrino:2016kp,kumar2017superballistic}, ${\eta}/{s} = T\nu$. The physical relevance of $\eta/s$ is that it yields the shear viscosity per effective degree of freedom participating in the momentum diffusion in the fluid. The ratio $\eta/s$ depends sensitively on $\alpha$, as shown in Fig.~\ref{fig:eta/s}. In the weakly Coulomb interacting regime, i.e. for $\alpha \ll 1$, first order perturbation theory such as Boltzmann's kinetic theory predicts a fast fall off, $\eta/s\sim \alpha^{-2}$, as shown by the black line in Fig.~\ref{fig:eta/s}. For intermediate Coulomb couplings, perturbative approaches lose their validity. Holographic gauge/gravity duality~\cite{Maldacena:1997re,Gubser:1998bc,Witten:1998qj} provides a nonperturbative approach to predict the coupling dependence of $\eta/s$. In the limit of infinitely strong coupling and for systems with a large number of degrees of freedom, it predicts the universal value~\cite{Policastro:2001yc,Buchel2004,Kovtun2005} (see the red dashed line in Fig.~\ref{fig:eta/s})
\begin{equation}\label{eq:etasAdSCFT}
\frac{\eta}{s} = \frac{1}{4\pi} \frac{\hbar}{k_B}\,.
\end{equation}
The essential feature of this result is that it is significantly smaller than any value of $\eta/s$ obtained within weak coupling perturbation theory.
Beyond the infinite coupling limit, gauge/gravity duality allows to include finite coupling corrections to the infinite coupling result
\begin{equation}\label{eq:FiniteCouplingCorrectionsMain}
{\eta \over s} = {\hbar \over 4\pi k_{\rm B}}\left(1 + \frac{\cal C}{\alpha^{3/2}}\right),
\end{equation}
where $\alpha$ is the fine-structure constant and the constant ${\cal C}$ parametrizes the class of gauge/gravity duals considered (see Methods).

\begin{figure*}[t!]
\centering
\includegraphics[width=\textwidth]{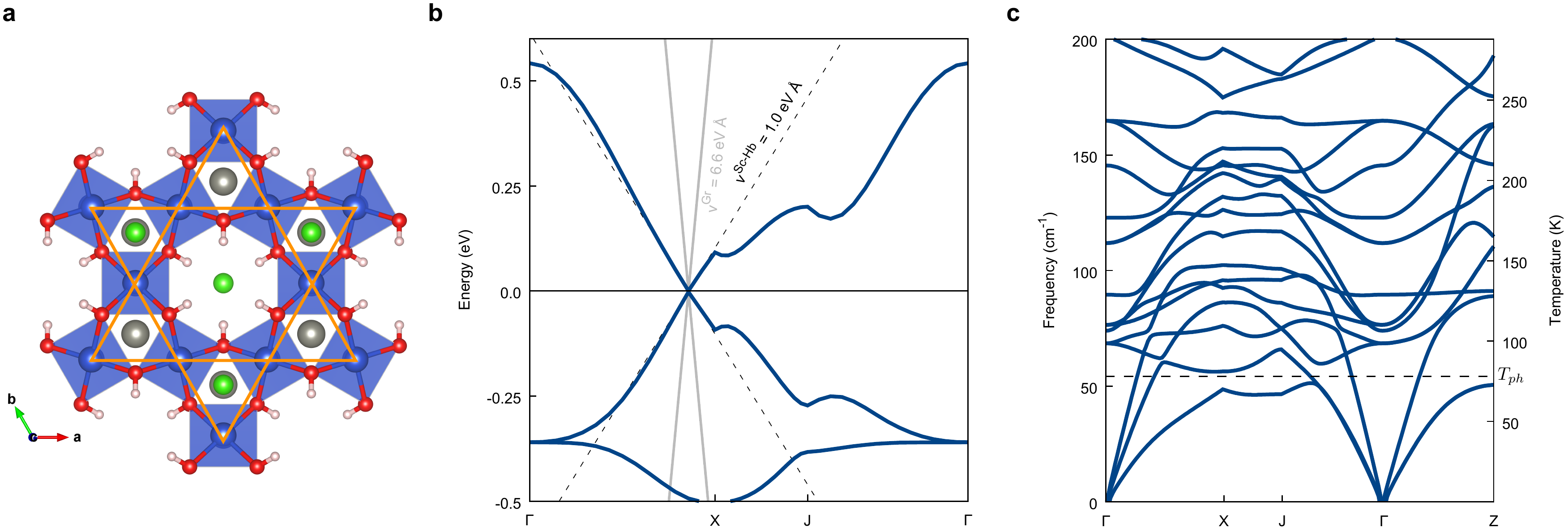}
\caption{
{\bf First-principles analysis of Sc-Herbertsmithite.} {\bf a}
Top view of the crystal structure, where the CuO$_4$ plaquettes
form a Kagome lattice, as highlighted by the orange lines. {\bf b}
Ab-initio band structure of the low-energy manifold of Sc-Herbertsmithite along the
high-symmetry directions of the conventional hexagonal Brillouin zone~\cite{sc-herbert}.
The dashed black lines refer to a linear fit around the Dirac point of
Sc-Herbertsmithite, whereas the solid grey lines denote the counterpart for graphene.
{\bf c} Phonon dispersion of Sc-Herbertsmithite. The horizontal dashed line marks the
temperature $T_{ph}$ above which optical phonon modes are thermally
activated.
}
\label{fig:Sc-Hb}
\end{figure*}

So far, the material of choice to investigate hydrodynamics of electronic systems has been graphene~\cite{novoselov2005two}. Here we show that in certain Kagome materials, hydrodynamic behaviour will be significantly enhanced.  Specifically, we focus on Kagome materials at filling levels such that the chemical potential is located at the Dirac point.  These materials are particularly suited for hydrodynamic studies since not only the Coulomb interaction is enhanced as compared to graphene (see below), but also because the Kagome lattices structure suppresses
the formation of ordered phases. The reason is that, in contrast to graphene, the Dirac cones on the Kagome lattice are located far away from half filling. As explained in Methods, combined with the small low energy density of states, this suggests a strong resilience of the metallic Dirac state against ordering instabilities~\cite{PhysRevLett.110.126405,PhysRevB.74.064429}, which implies that it sustains stronger Coulomb coupling than a Dirac metal on the honeycomb lattice~\cite{PhysRevX.3.031010}. For the explicit candidate Kagome metal Scandium-substituted Herbertsmithite ScCu$_3$(OH)$_6$Cl$_2$ (Sc-Herbertsmithite hereafter, see Fig.~\ref{fig:Sc-Hb})~\cite{sc-herbert}, we further calculated the phonon spectrum and find that the optical phonons decouple from the electronic degrees of freedom for temperatures below $\sim$80K, ensuring that a Kagome Dirac metal with electronic interactions is the appropriate microscopic description.

The suppression of gapped ordered phases and the decoupling of phonons in the proposed Kagome materials allows the electrons to form a fluid by Coulomb interactions up to very low temperatures. The electronic Dirac fluid of particle-hole excitations around the Dirac point then has emergent relativistic symmetry, is parity and time reversal invariant, and particle-hole symmetric. With these symmetries in place, a relativistic fluid is described by relativistic hydrodynamic equations of motion, which depends on the following key parameters: The Fermi velocity $v_F$ of the relativistic disperion relation playing the role of the speed of light, the relative dielectric constant $\epsilon_r$ in the medium, and the shear viscosity $\eta$ of the fluid. {As explained in Methods, the two other transport coefficients bulk viscosity $\xi$ and interaction induced conductivity $\sigma_Q$ are not relevant for our arguments.} The quantities $v_F$ and $\epsilon_r$ set the value of the effective Coulomb coupling $\alpha$ via 
\begin{equation}
\label{eq:alpha}
\alpha = \frac{e^2}{\epsilon_0 \epsilon_r \hbar v_{\rm F}}\,,
\end{equation}
where $\epsilon_0$ is the dielectric constant in vacuum. In turn, $\eta$ depends on $\alpha$. We calculate $v_F$ and $\epsilon_r$, and hence $\alpha$, within the framework of the constrained Random Phase Approximation (cRPA) (see Methods), and compare the results for Sc-Herbertsmithite with (hBN encapsulated) graphene. While our findings are likely to be applicable to a broad class of Kagome metals, Sc-Herbertsmithite suitably underlines a prime motif of how to accomplish Dirac fillings in Kagome systems. In pristine Herbertsmithite, Zn$^{2+}$ acts as a charge donor to the Cu d$_{x^2-y^2}$ orbitals which dominate the low energy theory, yielding a half filled Kagome lattice setting. Synthesis of the otherwise identical compound with Zn$^{2+}$ replaced by Sc$^{3+}$ provides the precise stoichiometry for the electrochemical potential to coincide with the Dirac points~\cite{sc-herbert}.

\begin{table}[b!]
\caption{
{\bf Dirac fluid parameters.} The Fermi velocity $v_F$, the relative dielectric constant $\epsilon_r$ and the fine-structure constant $\alpha$ for electrodynamics in vacuum, (hBN encapsulated) graphene~\cite{Roesner,Katsnelson12} and stochiometric Scandium substituted Herbertsmithite. In graphite ~\cite{Wehling} the low-energy dispersion at the K point of the Brillouin zone is
quadratic and not linear as in graphene.
}
\begin{center}
    \begin{tabular}{ | c | c | c | c |}
    \hline
                       & $v_F$ (eV\AA)  & $\epsilon_r$              & $\alpha=e^2/\epsilon_0 \epsilon_r\hbar v_F$ \\ \hline
    ED in vacuum       & $2\times 10^3$ & 1                       & 1/137                          \\ \hline
    hBN/graphene/hBN          & $6.6$          & 2.2 -- 4.0                 & 0.5 -- 1.0                            \\ \hline
    graphite          &        --       & 2.5                     &              --                  \\ \hline
    Sc-Herbertsmithite & $1.0$          & 5.0                     & 2.9                            \\
    \hline
    \end{tabular}
\end{center}
\label{tab:tab1}
\end{table}

Our first result (c.f.\ Table~\ref{tab:tab1}) is that the fine-structure constant $\alpha$ in Sc-Herbertsmithite is more than three times larger than in graphene, implying a strong enhancement in the applicability of viscous hydrodynamics in this material. With Sc-Herbertsmithite being our candidate material for the realization of holographic hydrodynamics, we refer to graphene as a benchmark for our theoretical analysis, as well as a reference point at weaker coupling, which helps us to identify expected trends in observables in more strongly coupled Dirac materials. 

Our second result is an estimate of the leading finite coupling correction in equation~\eqref{eq:FiniteCouplingCorrectionsMain} within a broad class of holographic models, as explained below and in Methods. This estimate leads to the blue band in Fig.~\ref{fig:eta/s}, allowing for predictions for the possible range of $\eta/s$ for both graphene and Sc-Herbertsmithite. Our predictions are indicated by the blue bars and show not only a considerably smaller value $\eta/s$ for Sc-Herbertsmithite as compared to graphene, but also a smaller variance. In particular, our holographic estimate shows that $\eta/s$ for the Dirac fluid in Sc-Herbertsmithite lies significantly closer to the infinite coupling value of equation~\eqref{eq:etasAdSCFT} predicted by  gauge/gravity duality, than in graphene. This makes the Dirac fluid in Sc-Herbertsmithite an interesting candidate for experimental realizations of holographic hydrodynamics. 

We can understand the difference in the scales for Sc-Herbertsmithite as compared to graphene along the following lines. In order for an electron system to display hydrodynamic behavior, the electron-electron mean free path must obey
\begin{equation} \label{eq:leeMain}
\ell_{\rm ee}(\alpha)\ll \ell_{\rm imp},\ell_{\rm ph},\,W,
\end{equation}
where $\ell_{\rm ph}$ is the electron-phonon mean free path, $\ell_{\rm imp}$ is the electron-impurity mean free path and $W$ the width of the sample. From equation~\eqref{eq:leeMain}, we infer that the emergence of hydrodynamic flow in solids is closely related to the relative value of the characteristic scales of the system~\cite{Molenkamp:1994ii,Molenkamp:1994kb}. This implies that the formation of viscous flow in solids is restricted to specific, and sometimes narrow, regimes. 
For Dirac fluids satisfying $\mu\ll k_B T$, the emergence of hydrodynamic behavior is greatly enhanced~\cite{polini2019viscous}, since in this large temperature regime the available phase space for electron-electron collisions is vastly enhanced. In turn, Landau quasiparticles are short lived, which yields a decrease of $\ell_{\rm ee}$. In addition, since $\ell_{\rm ee}\propto 1/\alpha^2$, increasing the Coulomb interaction strength leads to shorter thermalization length and time scales. This leads to an enhancement of hydrodynamic behaviour and a higher tendency to exhibit viscous flow~\cite{polini2019viscous}. In addition, the emergence of additional non-hydrodynamic modes sets a critical length scale $\ell_{\rm c}$, below which the standard hydrodynamic approach ceases to be valid. For the particular case of Sc-Herbertsmithite, we find (see Methods) 
\begin{equation}
\label{eq:Crit.q_ratio}\ell^{\rm Hb}_{\rm c}  \simeq  \frac{1}{6}\ell^{\rm Gr}_{\rm c}.
\end{equation}
This shows that the hydrodynamic regime is more robust in Sc-Herbertsmithite than in graphene.

Note that the shear viscosity controls the turbulent behaviour of the fluid, an aspect of electron hydrodynamics that still remains largely unexplored.  Within the hydrodynamic regime, the relative size of dissipative forces to nonlinear inertial forces in the fluid distinguishes the laminar flow regime, in which viscous forces dominate, from the turbulent flow regime, where non-linear dynamics in the Navier-Stokes equations dominate. More specifically, the ratio of inertial to dissipative/viscous forces is characterized by the Reynolds number of the flow. For a $2+1$ dimensional Dirac fluid at charge neutrality flowing within a channel of width $W$, the Reynolds number is given by
\begin{equation}
\label{eq:Reynolds-number}Re = \left( {\eta \over s}{k_{\rm B} \over \hbar}\right)^{-1} {k_{\rm B} T\over \hbar v_{\rm F}} {u_{\rm typ}(\eta/s) \over v_{\rm F}} W,
\end{equation} 
where  $s$ the entropy density of the fluid and $u_{\rm typ}$ is the typical velocity of the fluid, which strongly  increases as $\eta/s$ decreases~\cite{Erdmenger:2018svl}. The fluid exhibits turbulent behaviour when $Re = {\cal O}(1000)$~\cite{doi:10.1146/annurev-fluid-010719-060221} .

The onset of turbulence is related to the fluid flow, which is sensitive not only to the sample geometry, but also to the intrinsic transport properties such as its shear viscosity---from equation~\eqref{eq:Reynolds-number}, $ Re \propto (\eta/s)^{-1} u_{\rm typ}(\eta/s)$. Since the shear viscosity is much smaller for strongly coupled fluids (see Fig.~\ref{fig:eta/s}), the experimental realization of turbulence will be greatly enhanced for materials exhibiting large Coulomb interactions. This conclusion holds regardless of the particular geometry employed. 
Furthermore, even laminar flows such as the Poiseuille channel flow will, compared to graphene or other even more weakly coupled materials, exhibit enhanced hydrodynamic transport properties such as larger typical fluid velocities and smaller differential resistances~\cite{Erdmenger:2018svl}.

The third and final result of this work is an estimate of the Reynolds number for the flow of our proposed holographic Dirac fluid in Sc-Herbertsmithite through a long and straight channel flow as described by equation~\eqref{eq:Reynolds-number}. In graphene, the Reynolds number has been found to be sufficiently large for preturbulent behavior such as vortex production, but not for fully developed turbulence~\cite{mendoza2011preturbulent}. For Sc-Herbertsmithite, we find an enhancement of the Reynolds number compared to graphene by a factor of order 100, mostly due to the six times smaller Fermi velocity $v_F$ (see Fig.~\ref{fig:Sc-Hb}{\bf b} and Table~\ref{tab:tab1}, as well as Methods for the estimate), leading to Reynolds numbers sufficiently large for fully turbulent channel flows. We hence predict that Sc-Herbertsmithite brings, for the first time, the turbulent flow regime in electron hydrodynamics within experimental reach.


\vspace{0.5cm}

\noindent {\bf Acknowledgements}\\
We thank Andy Lucas for useful discussions. The work in W\"urzburg is funded by the Deutsche Forschungsgemeinschaft (DFG, German Research Foundation) through Project-ID 258499086 - SFB 1170 and through the W\"urzburg-Dresden Cluster of Excellence on Complexity and Topology in Quantum Matter --\textit{ct.qmat} Project-ID 39085490 - EXC 2147. We acknowledge the Gauss Centre for Supercomputing e.V.  for funding this project by providing computing time on the GCS Supercomputer SuperMUC at Leibniz Supercomputing Centre.

\noindent {\bf Author contributions}\\
J.E., M.G., R.M., and R.T.\ initialized the project.  J.E.\ and R.T.\ coordinated and supervised the investigation. I.M., R.M., and D.R.F.\ developed and performed the AdS-CFT analysis, while D.D.S., E.v.L., and T.W.\ performed the ab initio and constrained RPA calculations.  I.M., D.D.S., J.E., M.G., R.M., D.R.F., and R.T.\ wrote the paper. The manuscript reflects contributions from all authors.

\noindent {\bf Competing interests}\\
The authors declare no competing interests.

\noindent {\bf Correspondence}\\
Correspondence and requests for materials should be addressed to R.T.~(email: rthomale@physik.uni-wuerzburg.de).

\vspace{0.5cm}

\noindent {\bf Methods}\\
\noindent {\bf Properties of Herbertsmithite.}\label{sec:Herbert} Kagome materials such as
Herbertsmithite combine the features of Dirac fermions and strong correlations. ScCu$_3$(OH)$_6$Cl$_2$, which we refer to as Sc-Herbertsmithite, consists of CuO$_4$ plaquettes forming a Kagome lattice (see Fig.~\ref{fig:Sc-Hb}{\bf a}). The crystal field is such that the low-energy physics is correctly captured by a single $d_{x^2-y^2}$ orbital on each Cu site, and the resulting low-energy band structure is qualitatively similar to that of a one-orbital model at $n=4/3$ electron filling, where, as shown in Fig.~\ref{fig:Sc-Hb}{\bf b}, the Fermi level is pinned at the Dirac points.

As compared to graphene, where the underlying Dirac spectrum originates from weakly correlated $p_z$ orbitals, Sc-Herbertsmithite is expected to show a larger degree of electronic correlations. This is confirmed by our constrained random phase approximation (cRPA) estimates of the coupling strength $\alpha$. For a linear band dispersion, the strength of the Coulomb interaction is characterized by the effective fine-structure constant in equation~\eqref{eq:alpha}.  According to the values summarized in Table~\ref{tab:tab1}, $\alpha^\text{Sc-Hb} = 2.9$ as compared to $\alpha^\text{Gr} = 0.5 - 1.0$ for (hBN-encapsulated) graphene. We use $\alpha = 0.9$ as a representative example in the main text..  The latter value was evaluated using an ab-initio estimate of the dielectric constant~\cite{Roesner,Wehling}.

\noindent {\bf First principle calculations.}\label{sec:Interaction} For
our numerical study of Sc-Herbertsmithite, we employed state-of-the-art first-principle calculations based on the density functional theory as implemented in the Vienna ab initio simulation package (VASP)~\cite{PhysRevB.54.11169} following the projector-augmented-plane-wave (PAW) method~\cite{PhysRevB.59.1758,PhysRevB.50.17953}. We use the generalized gradient approximation as parametrized by the PBE-GGA functional for the exchange-correlation potential~\cite{PhysRevLett.77.3865} by expanding the Kohn-Sham wave functions into plane waves up to an energy cutoff of 400 eV and sampling the Brillouin zone on an 6$\times$6$\times$6 regular mesh. We obtain phonon dispersions within the context of the frozen phonon method~\cite{PhysRevLett.78.4063} as implemented in the PHONOPY code~\cite{PhysRevB.78.134106} with a 2$\times$2$\times$2 supercell. We subsequently determine the Fermi velocity by a fit to the band structure along the $\Gamma$-X-J-$\Gamma$ path. The effective dielectric constant, which describes the screening of the Cu $d_{x^2-y^2}$ electrons by the other electrons further away from the Fermi level, is calculated from the dielectric tensor. We fixed (constrained) the occupation of all electronic states in the three Kagome bands to be $n = 4/3$ (including a factor of 2 from spin degeneracy), to exclude intraband screening processes~\cite{cRPA}. To a first approximation, the dielectric tensor is diagonal and given by a single constant $\epsilon_r$.

\noindent {\bf Phonons.} A number of caveats need to be addressed.  Interactions of the electron fluid with lattice vibrations are detrimental to the hydrodynamic behaviour. As shown in Fig.~\ref{fig:Sc-Hb}{\bf c}, however, the optical phonon modes in Sc-Herbertsmithite are populated for temperatures above $T_{\text{ph}}\sim 80$K. This analysis indicates that the hydrodynamic regime can be expected to extend over a range of temperatures within current experimental reach.

\noindent {\bf Instabilities.}\label{sec:instabilities} The Kagome lattice itself plays a crucial role. In contrast to graphene's Dirac spectrum, which occurs precisely at half-filling, and as such it is prone to magnetic instabilities at comparably small couplings~\cite{PhysRevX.3.031010}, the Dirac fluid in Kagome Sc-Herbertsmithite is reached at the stoichiometric filling $n=4/3$. Therefore, tendencies towards a correlation driven magnetic groundstate are strongly suppressed, and the linear dispersion of the Dirac fluid is expected to be highly robust against the opening of a band gap. While the geometric frustration inherent in the Kagome lattice
precludes a rigorous treatment, this conjecture is supported both from weak coupling~\cite{PhysRevB.86.121105,PhysRevLett.110.126405} as well as strong coupling studies~\cite{PhysRevB.74.064429}.

\noindent {\bf Hydrodynamics.} Electronic fluids are in local thermal equilibrium on length scales larger than the interaction mean free path $\ell_{\text{ee}}$, which needs to be the smallest length scale for hydrodynamics to apply (see equation~\eqref{eq:leeMain}). Locally, the laws of thermodynamics such as the Gibbs-Duhem relation
\begin{equation}
\label{eq:GibbsDuhem} \epsilon+P = s T + \mu \rho
\end{equation}
holds for the densities of energy $\epsilon(T,\mu)$, entropy $s(T,\mu)$ and charge $\rho(T,\mu)$. Given the equation of state $P(\epsilon,\rho)$, equation~\eqref{eq:GibbsDuhem} can be used to determine the entropy density $s(\mu,T)$. A two-dimensional Dirac fluid, i.e. a fluid of electrons with relativistic dispersion and the chemical potential pinned to the Dirac point, $\mu=0$, is then described by relativistic hydrodynamics~\cite{Landau:1987}. The hydrodynamic equations are the conservation equations of the energy-momentum tensor $T^{\mu\nu}$ and the electric current $J^\mu$,
\begin{equation}\label{relhydroeqs}
\partial_\mu T^{\mu\nu} =  F^{\nu\mu}J_\mu,\
\partial_\mu J^\mu = 0,
\end{equation}
where we have also allowed for the coupling of the fluid to an external electromagnetic field, $F^{\nu\mu}$.

The hydrodynamic derivative expansion expresses $T^{\mu\nu}$ and $J^\mu$ order by order in derivatives of the local temperature $T(x^\nu)$, the chemical potential $\mu(x^\nu)$, and the relativistic 3-velocity $u^\mu(x^\nu)$ ($u^\mu u_\mu = -c^2$). For a parity and time reversal invariant, relativistic two-dimensional fluid, one obtains to first order in the derivatives
\bea
T^{\mu\nu} &=& T^{\mu\nu}_{\rm (0)} + T^{\mu\nu}_{\rm (1)},\\ \label{eq:T2}
T^{\mu\nu}_{\rm (0)} &=& \varepsilon\, u^\mu u^\nu/c^2 + p \Delta^{\mu\nu}\,, \\ \label{eq:T3}
T^{\mu\nu}_{\rm (1)} &=& -\eta \Delta^{\mu\alpha}\Delta^{\nu\beta}\left(2\partial_{(\alpha} u_{\beta)}-\Delta^{\alpha\beta} \partial_\sigma u^\sigma \right) \nonumber\\
&\quad& -\xi \Delta^{\mu\nu} \partial_\gamma u^\gamma ,\\
\label{eq:j}
J^\mu &=& e\rho u^\mu + \sigma_Q \left(E^\mu - T \Delta_{\mu\nu} \partial^\nu \frac{(\mu/e)}{T}\right),
\eea
where $\Delta^{\mu\nu} = u^\mu u^\nu/ c^2 + \eta^{\mu\nu}$ is the projection matrix in the directions transverse to $u^\mu$ and $A_{(\mu\nu)} = (A_{\mu\nu} + A_{\nu\mu})/2$ denotes symmetrization of any two-tensor.


The shear viscosity $\eta$ is the main observable in this context. It can be calculated from kinetic theory in the perturbative regime $\alpha \ll 1$~\cite{Muller:2009cy}, and will be derived from a holographic model below in the nonperturbative regime $\alpha\gg 1$. The bulk viscosity $\xi$ is negligible due to the approximate scale invariance of the linear dispersion relation in a Dirac fluid, $\xi\approx 0$. For incompressible flows ($\partial_\mu u^\mu = 0$) such as the Poiseuille flow, a finite bulk viscosity has no effect either. The interaction-induced intrinsic conductivity $\sigma_Q$~\cite{Fritz:2008go} affects electric transport and controls the rate of Joule heating, but does not affect the flow of the fluid.

The three transport coefficients $(\eta,\xi,\sigma_Q)$ have to be calculated from a microscopic theory or an effective model, such as the kinetic theory at weak coupling or holography at strong coupling. In linear response theory around a global equilibrium (i.e.\ constant temperature $T$ and chemical potential $\mu$, and $u^\mu=(1,0,0)^T$), the shear viscosity $\eta$ can be calculated with the following Kubo formula at zero frequency $\omega$ and momentum $k$,
\begin{equation}
\label{eq:Strong-a-ratio} {\eta} = \lim_{\omega \rightarrow 0} {1 \over 2i\omega}  \langle[T_{\rm xy}(\omega), T_{\rm xy}(0)] \rangle .
\end{equation}
In the strong coupling regime, the AdS/CFT correspondence provides a framework to calculate $\eta$ via a semi-classical analysis of the dual gravitational action of the system~\cite{Cremonini:2011iq,Buchel:2008vz}.  The entropy density of the fluid is then given by the entropy density of the black hole horizon in the bulk of spacetime~\cite{GUBSER1998202}.

\noindent {\bf Holographic Model.} At infinite coupling and for a large number of degrees of freedom, the holographic dual is given by the Einstein-Hilbert action
\begin{equation}
\label{eq:EHAction} 
S_{\text{EH}} = \frac{1}{16\pi G_N} \int d^4 x \sqrt{-g} \left (R - 2 \Lambda\right),
\end{equation}
yielding equation~\eqref{eq:etasAdSCFT} for the ratio $\eta/s$, independently of the coupling constant and number of degrees of freedom. Our corrections to this limit are computed by adding higher powers of the curvature $R$ to equation~\eqref{eq:EHAction}. The next-to-leading order $R^2$ terms contribute a topological Gauss-Bonnet term to the gravitational action \cite{Brigante:2007nu} and therefore do not alter the value of $\eta/s$. {It was shown in \cite{Metsaev:1986yb} that type II supergravity, the ten-dimensional parent theory of our four-dimensional gravity dual, does not contain $R^3$ corrections. The next higher derivative corrections contain four powers of $R$~\cite{Buchel:2004di,Benincasa:2005qc,Policastro:2001yc} and are induced in type II supergravity by quartic terms involving the Weyl tensor.} These terms yield 
\begin{equation}\label{eq:FiniteCouplingCorrections}
\label{eq:eta over s} {\eta \over s} = {\hbar \over 4\pi k_{\rm B}}\left(1 + \frac{\cal C '}{\lambda^{3/2}}\right).
\end{equation}
In top-down constructions of holography originating from string theory, the holographic dual theory is typically a non-Abelian gauge theory with gauge group rank $N$ and 't~Hooft coupling $\lambda = \alpha N$, related to the fine-structure constant $\alpha$. $N$ parametrizes the number of degrees of freedom in the dual gauge theory.  {Dimensional analysis implies that the correction at order $R^4$ to $\eta/s$ scales as $\lambda^{-3/2}$ in any spacetime dimension, up to a multiplicative constant that depends on the details of the holographic model  \cite{Cremonini:2011iq}. }Equation~\eqref{eq:FiniteCouplingCorrections} {in particular} universally describes the coupling dependence of the ratio $\eta/s$ for all  $(3+1) d$-holographic duals with relativistic symmetry at leading order in the inverse coupling expansion and in the large $N$ limit near charge neutrality. However, the prefactors of the allowed subleading curvature corrections are  model dependent. We parametrize the model dependence of the $R^4$ correction through the prefactor ${\cal C '}$ in equation~\eqref{eq:FiniteCouplingCorrections}. {For the original example of a theory with holographic dual, $\mathcal{N} = 4$ SYM theory \cite{Maldacena:1997re}, ${\cal C '}= 135\zeta(3)/8$ \cite{Buchel:2004di}.} The unknowns in equation~\eqref{eq:FiniteCouplingCorrections} are $N$ and ${\cal C '}$, with ${\cal C '}$ depending on the particular holographic dual  considered.  We absorb $N^{-3/2}$ into ${\cal C '}$ via ${\cal C} \equiv {\cal C'}N^{-3/2}$, which brings equation~\eqref{eq:FiniteCouplingCorrections} to the form of equation~\eqref{eq:FiniteCouplingCorrectionsMain}. Thus, the details of the holographic model suitable for describing Dirac fluids are encoded in a single coefficient, namely ${\cal C}$. {Using the parametrization in terms of ${\cal C}$,} we generate the blue band in Fig.~\ref{fig:eta/s} as follows: We vary ${\cal C}$ from ${\cal C} = 0.0005$ to ${\cal C} = 5$ to generate the blue band in Fig. 1. {For $\mathcal{N} = 4$ SYM, this range of variation of ${\cal C }$ corresponds to formally varying $N$ from $N \simeq 10^3$ to $N \simeq 2$.}
As shown in Fig 1, even varying ${\cal C}$ over four orders of magnitude changes the value of $\eta/s$ of Sc-Hb by at most a factor of two. In principle, there may be corrections of even higher order in $1/\lambda$  to equation~\eqref{eq:FiniteCouplingCorrections}, corresponding to terms involving even higher orders in the curvature. However 
we do not expect these to alter the predicted estimate of the value of $\eta/s$  for values of the coupling not accessible to weak-coupling perturbation theory. Similarly, there may be $1/N$ corrections independent of $\alpha$, however in agreement with  \cite{Kats:2007mq} we expect these to be subleading as compared to the ${\cal C}/\alpha^{3/2}$ corrections at  $\alpha^\text{Sc-Hb} = 2.9$, for the range of ${\cal C}$ chosen. The black, fading, line in Fig.~\ref{fig:eta/s} shows the perturbative prediction at weak coupling~\cite{Fritz:2008go,Muller:2009cy}, which cannot be extrapolated to stronger coupling without violating the condition $\eta\geq 0$ that follows from local application of the second law of thermodynamics.

\noindent {\bf Turbulence.} Using the extrapolation presented in Fig.~\ref{fig:eta/s}, we infer that materials which display a stronger Coulomb interaction, such as Sc-Herbertsmithite, yield a much larger Reynolds number than that reported for graphene. In particular, using equation~\eqref{eq:Reynolds-number},  Fig.~\ref{fig:eta/s} and the results for the Fermi velocity $v_F$ from Table~\ref{tab:tab1}, we find that Sc-Herbertsmithite will exhibit a $60$ to $156$ times larger Reynolds number as compared to graphene, depending on whether we use the value of $\eta/s$  at the bottom or the top of the blue band, respectively. This enhancement is large enough to expect fully developed turbulence in a constriction setup \cite{mendoza2011preturbulent}.

\noindent {\bf Coupling dependence of the regime of validity of hydrodynamics.}\label{sec:validity}
In this section, we derive equation~\eqref{eq:Crit.q_ratio} for the relative ratio between the length scales where hydrodynamics is expected to break down for Sc-Herbertsmithite and graphene. One can characterize the breakdown of hydrodynamics by the existence of a diffusive pole, $ \omega = -i D k^2$, in the energy-momentum tensor self-correlation in equation~\eqref{eq:Strong-a-ratio}. Through the calculation of the poles of the mentioned self-correlation function via holography, the diffusive pole was shown~\cite{grozdanov2016strong} to disappear at a critical wavelength $\ell_{\rm c}$ as the coupling strength of the system is decreased.  The pole moves down the imaginary frequency axis and collides with an upward moving pole at $ \ell = \ell_{\rm  c}$. After the collision, the two poles split into a conjugate pair and the purely imaginary diffusion pole disappears. An approximate analytic result for $l_{\rm c}$ is given by~\cite{grozdanov2016strong} 
\begin{equation}
\label{eq:Crit.q} 
\ell_{\rm c} \simeq \frac{h v_F}{0.04 k_{\rm B}T} \lambda^{-3/2},
\end{equation}
where $\lambda$ is the 't Hooft coupling defined below equation~\eqref{eq:eta over s}. {Within the model employed~\cite{grozdanov2016strong}, the analytical approximation in equation~\eqref{eq:Crit.q} lies closer to the numerical value of $\ell_{\rm c}$ for smaller values of $\lambda$. To our knowledge, $\ell_{\rm c}$ was  derived only through holographic methods. At weak coupling, it is only known that hydrodynamics has a finite radius of convergence.} For the particular case of graphene and Sc-Herbertsmithite, with $\alpha^\text{Gr} = 0.9$ and $\alpha^\text{Sc-Hb} = 2.9$, we find the ratio given in equation~\eqref{eq:Crit.q_ratio}.
Note that we calculated only the ratio of critical wavelengths and not their absolute values. We did so because the two materials have the same low-lying energy spectra and thus equation~\eqref{eq:Crit.q_ratio} is independent of the unknowns $N, {\cal C}$ of equation~\eqref{eq:eta over s}. Equation~\eqref{eq:Crit.q_ratio} shows that if both graphene and Sc-Herbertsmithite are described by holography, then the hydrodynamic approximation is more robust for Sc-Herbertsmithite.  
This argument presumes that graphene can be described holographically. Because of  graphene's intermediate coupling strength, however, $\alpha^\text{Gr} = 0.9$, it is more likely to lie in the intermediate regime, where neither simple truncated holographic models nor perturbative methods are applicable. In this case equation~\eqref{eq:Crit.q_ratio} entails that Sc-based Herbertsmithite is a better candidate for the realization of holographic hydrodynamics.

\end{document}